\documentclass[journal, twoside]{IEEEtran}

\newcommand{\mytitle}{A Reanalysis of the October 2016 ``Meteotsunami'' in British Columbia with Help of High-Frequency Radars and Autoregressive Modeling}

\usepackage[
	hidelinks, pdftex,
	pdftitle={\mytitle},
	]{hyperref}
\title{\mytitle}

\newcommand{\trait}[1]{{\bfseries\color{#1} --------}}
\newcommand{\dottrait}[1]{{\bfseries\color{#1} -\,-\,-\,-\,-\,-}}

\newcommand{\dotligne}[2]{{\bfseries\color{#1} -\,-\,-\,-\,-\,-}~#2}
\newcommand{\ligne}[2]{{\bfseries\color{#1} --------}~#2}

\usepackage[pdftex]{graphicx}
\usepackage[svgnames]{xcolor}
\graphicspath{{figures/}}
\usepackage{transparent}
\usepackage{tikz}
\usepackage{dblfloatfix}    

\usepackage{amsmath}
\usepackage{amssymb}
\usepackage{amsfonts}
\usepackage{gensymb}
\usepackage{siunitx}
\usepackage{physics}
\renewcommand{\epsilon}{\varepsilon}
\usepackage{interval}

\renewcommand{\mathbb}[1]{\mathbf{#1}}
\everymath{\displaystyle}
\DeclareMathOperator{\Var}{Var}

\usepackage{multirow}
\usepackage{booktabs,tabularx}
\usepackage{array}
\usepackage[noadjust]{cite}
\usepackage{url}
\usepackage{enumitem}

\usepackage{soul}
\sethlcolor{yellow}

\author{Baptiste Domps, \IEEEmembership{Student Member, IEEE}, Julien Marmain, and Charles-Antoine Gu\'erin
	\thanks{Manuscript received XXXXXXXXX; revised XXXXXXXX; accepted XXXXXXXX. The work of Baptiste Domps was supported by the Direction G\'en\'erale de l'Armement (DGA) via the Agence pour l'Innovation de D\'efense (AID). \textit{(Corresponding author: Baptiste Domps.)}}
	\thanks{B. Domps and J. Marmain are with the Radar \& Scientific Applications Department, Degreane Horizon, 83390 Cuers, France (e-mail: baptiste.domps@degreane-horizon.fr; julien.marmain@degreane-horizon.fr).}
	\thanks{C.-A. Gu\'erin is with the Mediterranean Institute of Oceanography (MIO), Universit\'e de Toulon, Aix-Marseille University, CNRS, IRD, Toulon, France (e-mail: guerin@univ-tln.fr).}
	\thanks{Color versions of one or more of the figures in this letter are available online at \url{http://ieeexplore.ieee.org}.}
	\thanks{Digital Object Identifier XXXXXXXXX}
}

\IEEEpubid{\begin{minipage}{\textwidth}\ \\[12pt] \centering
  1545-598X \copyright~2021 IEEE. Personal use is permitted, but republication/redistribution requires IEEE permission.\\
  See \url{http://www.ieee.org/publications standards/publications/rights/index.html} for more information.
\end{minipage}}

\begin{document}

\maketitle

\begin{abstract} 
  On October 14, 2016, the  coastal high-frequency radar system in Tofino (British Columbia, Canada) triggered an automatic tsunami warning based on the identification of abnormal surface current patterns. This occurred in the absence of any reported seismic event but coincided with a strong atmospheric perturbation, which qualified the event as meteotsunami. We re-analyze this case in the light of a new radar signal processing method which was designed recently for inverting fast-varying sea surface currents from the complex voltage time series received on the antennas. This method, based on an autoregressive modeling combined with a maximum entropy method, yields a dramatic improvement in both the Signal-to-Noise Ratio and the quality of the surface current estimation for very short integration time. This makes it possible to evidence the propagation of a sharp wave front of surface current during the event and to map its magnitude and arrival time over the radar coverage. We show that the amplitude and speed of the inferred residual current do not comply with a Proudman resonance mechanism but are consistent with the propagation of a low-pressure atmospheric front. This supports the hypothesis of a storm surge rather than a true meteotsunami to explain this event. Beyond this specific case, another outcome of the analysis is the promising use of HF radars as proxy's for the characterization of atmospheric fronts.
\end{abstract}


\IEEEpeerreviewmaketitle

\section{Introduction}

\begin{figure}[h]
	\centering
	\includegraphics[width=\linewidth]{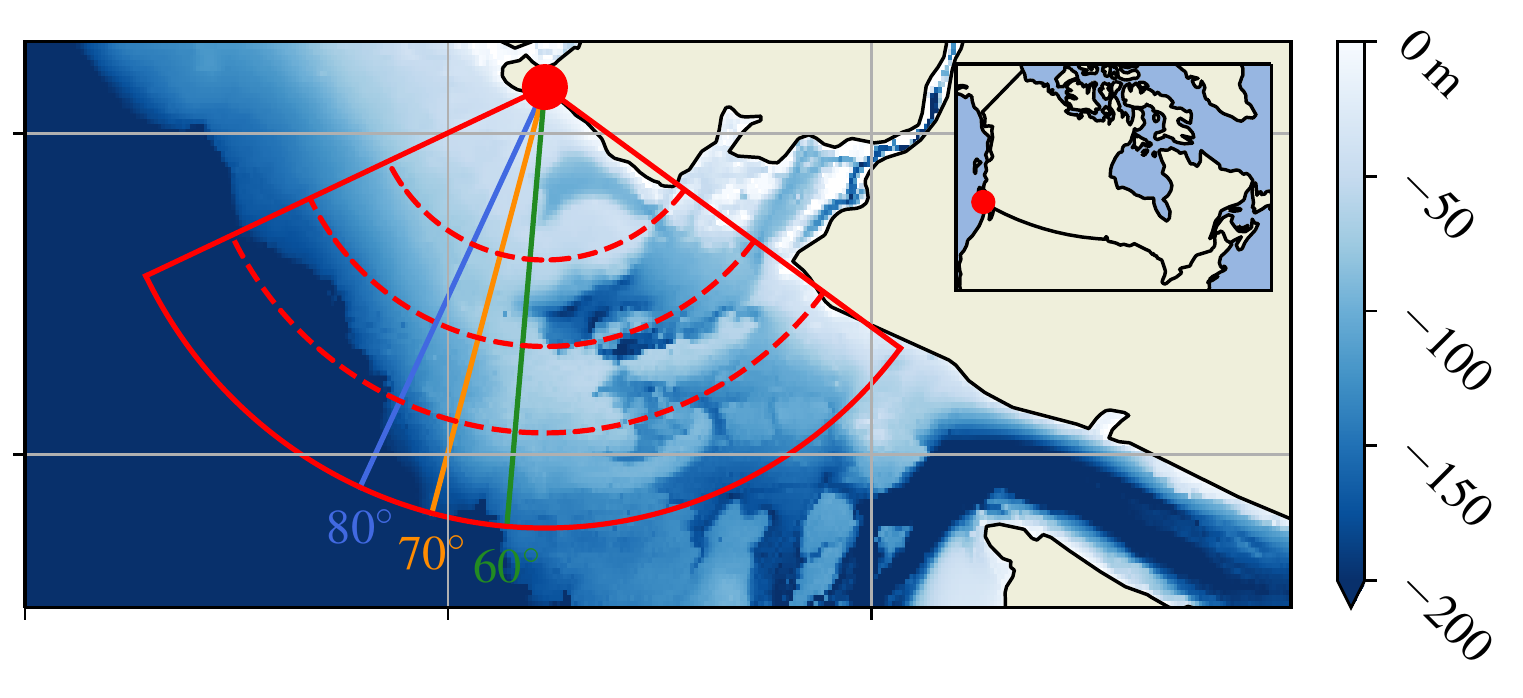}
	\caption{Bathymetry (colorscale; \si{\km}; limited to \SI{-200}{\m}), sweep area (\trait{red}\,; assuming a \SI{85}{\km} maximum range) and isorange contours 30, 45 and \SI{60}{\km} (\dottrait{red}) of the WERA HFR located in Tofino ({\color{red} $\bullet$}) along with bearings (W to E): \ligne{RoyalBlue}{\SI{80}{\degree}}, \ligne{DarkOrange}{\SI{70}{\degree}} and \ligne{ForestGreen}{\SI{60}{\degree}}.}.
	\label{fig:tofino}
\end{figure}

\IEEEPARstart{T}{sunami} early warning is one emerging application of high-frequency radars (HFR) systems. The concept was first proposed four decades ago \cite{barrick_RSE79} but it is only after the Indonesia 2004 and Tohoku Japan 2011 big tsunamis that it was actually confronted with real data (e.g. \cite{lipa2011japan,dzvonsko_IRS11,lipa2012tsunami,werachili2014}). Today, several HFR systems are equipped with a tsunami detection software in addition to their routine task of coastal current monitoring. One such instrument has been installed in 2015 for Ocean Networks Canada (ONC) in Tofino, on the West coast of Vancouver Island, British Columbia (Fig. \ref{fig:tofino}). It is a WERA system developed by Helzel Messetechnik GmbH operating at \SI{13.5}{\mega\hertz} and providing oceanographic measurements up to \SI{110}{\kilo\metre} to the South within a \SI{120}{\degree} sweep area.

\IEEEpubidadjcol

On October 14, 2016, at 06:06~UTC, the HFR of Tofino issued an automatic warning of high probability of tsunami \cite{aes:dzvonkovskaya2018}, which was the first ever in the short history of tsunami radar warning. Even though no seismic activity was reported at that time, the event was confirmed by the measurement of anomalous long-period sea level oscillations of about \SI{20}{\centi\metre} amplitude by a tide gauge in Tofino. This occurred in the context of a series of strong atmospheric low-pressure disturbances which hit British Columbia from October 13 to 16 \cite{irs:dzvonkovskaya2017, od:guerin2018} in the remnants of typhoon Songda and led to interpret this phenomenon as a meteorological tsunami \cite{osm:rabinovich2018}. Analysis of the recorded radar data confirmed the occurrence of abnormal residual current patterns in the form of a marked propagating ``jump'' in amplitude whose celerity was found consistent with the propagation speed of the low-pressure front which could be coarsely estimated from weather buoys \cite{od:guerin2018}. However, the exact physical mechanism at the origin of the anomalous residual waves and currents could not be definitely established and left open to several possibilities such as a meteotsunami with Proudman resonance or a mere storm surge.

In this letter we take advantage of an improved radar signal processing method which was applied recently in the context of HFR surface current retrieval \cite{joe:domps2020}. It is based on an autoregressive (AR) modeling  of the voltage time series combined with a maximum entropy method (MEM) for the estimation of the AR coefficients. While the surface current is classically derived from a Doppler analysis of the recorded signal, this non-spectral method allows to bypass the time-frequency dilemma and to obtain reliable estimates at very short-integration time, which is a prerequisite for tsunami detection \cite{heron_ocean15} (Section \ref{sec:tvar}). The inspection of residual current at high temporal rate reveals the propagation of a steep jump in amplitude which we interpret as the instantaneous response of the sea surface to the local atmospheric disturbance (Section \ref{sec:currents}). Using a change point detection method, the exact times of arrival and magnitude of the wave front of surface current can be accurately determined and charted (Section \ref{sec:cpd}). The joint analysis of the celerity (derived from the times of arrivals) and amplification of the residual current along the main propagation line disqualifies the Proudman resonance as the origin of the alert while the satellite imagery of GOES-15 favors the hypothesis of a storm surge (Section \ref{sec:discussion}).

\section{Time-Varying Autoregressive Modeling of the Doppler Oceanic Spectrum}\label{sec:tvar}

\begin{figure*}[!b]
	\centering
	\includegraphics[width=.95\linewidth]{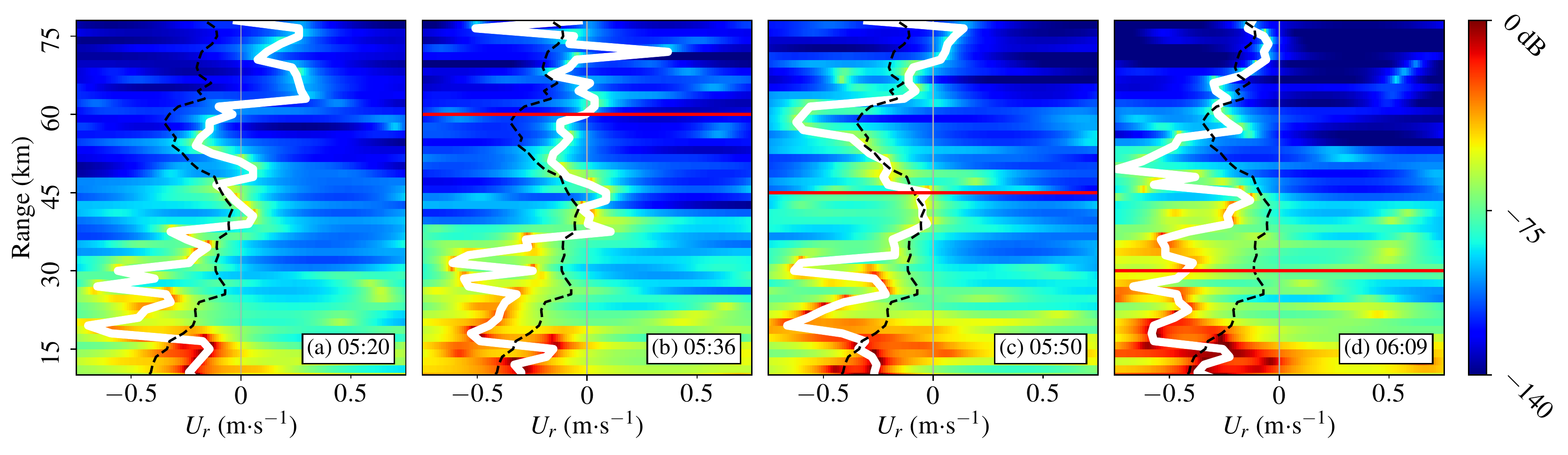}
	\caption{High-resolution Range-Doppler power spectra (\si{\deci\bel}; colorscale) centered on the negative Bragg line $-f_B$ (i.e. zero-Doppler for the surface current) and computed with the AR-MEM from samples of $N=$~\SI{128}{points} (\SI{33}{\s}) along azimuth \SI{70}{\degree} on October 14, 2016. The propagation of a front in current magnitude is marked by \trait{red} at: (a)~05:20 (phenomenon out of range); (b)~05:36; (c)~05:50; (d)~06:09~UTC. This front corresponds to a shift of $U_r$ (thick white line) with respect to the \dotligne{black}{mean ``background'' oceanic current}, from positive to negative radial speeds.}
	\label{fig:rd_meteo}
\end{figure*}

Today, high-frequency (HF) radars are routinely used for the monitoring of coastal surface currents \cite{front:roarty2019}. The measurement is based on the evaluation of the sea surface radar cross-section per unit bandwidth  $\sigma(f)$, most often simply referred to as the Power Spectrum Density (PSD) or the backscattered ``Doppler spectrum''. As it has been well known since the pioneering works of Crombie \cite{nat:crombie1955} and Barrick \cite{tap:barrick1972}, the most salient feature of the Doppler spectrum is a pair of marked spectral rays located at the so-called Bragg resonant frequencies $\pm f_B$, which are given in the absence of surface current by $f_B=\sqrt{\pi/(g\lambda_0)}$, with $g$ being the standard gravity and $\lambda_0$ the radar wavelength. In the presence of a surface current, the Doppler spectrum is shifted by an additional frequency $f_c$ which corresponds to the celerity $U_r$ of the current in the radar look direction (the so-called radial speed), $f_c=2U_r/\lambda_0$. A mapping of the radial surface current is achieved by evaluating the actual position of the Bragg rays for each radar cell and calculating the induced frequency shift with respect to the theoretical value $\pm f_B$. The on-board computation of the Doppler spectrum is in the vast majority of cases performed using a Fast Fourier Transform (FFT) of the recorded backscattered time series. This is by far the most efficient numerical method and is amply satisfactory in most operational situations. However, as it is well known, the Fourier analysis is bound to a time-frequency trade-off which prevents from using short integration times as this would deteriorate drastically both the accuracy and the Signal to Noise Ratio of the Bragg peak estimation. Some alternative, non-spectral methods have therefore been proposed in the literature to cope with the necessity of short observation windows to monitor fast-varying physical phenomena (e.g. \cite{om:guerin2018,od:guerin2018}). Very recently, it was shown \cite{joe:domps2020} that the use of a parametric approach based on autoregressive (AR) modeling of the backscattered time series is very promising in addressing this issue and well-performing for integration time as short as one minute. In the AR approach, the instantaneous received complex signal at the sample rate $\Delta t$ is expressed as a linear combination of the previous values in the past together with an additive white noise $\varepsilon[n]$:
\begin{equation}
	\label{eq:ar}
	s(n\Delta t) = -\sum_{k=1}^pa[k]s\big((n-k)\Delta t\big)+\varepsilon[n]
\end{equation}
The number $p$ of involved values in the past is called the order of the AR model and the parameters $a[k]$ are the AR coefficients, which can be estimated using different schemes (see, e.g., \cite{book:stoica2005}). In a previous work \cite{joe:domps2020}, the authors assessed the estimation and performances of the AR model for the estimation of surface current. For very short samples, the best performing method for determining the AR coefficients was found to be the so-called maximum entropy method (MEM), sometimes better known as the ``Burg method'' \cite{phd:burg1975}. In the case of a radar signal associated to a stationary surface current with non-varying AR coefficients, it was verified on synthetic numerical test cases that the optimal choice of the AR order $p$ for a time series of length $N$ is about $N/2$. Once the AR coefficients have been estimated, the corresponding PSD is obtained from (e.g. \cite{book:marple2019}):
\begin{equation}
	\label{eq:ar_psd}
	P_{AR}(f) = P_\varepsilon\left|1+\sum_{k=1}^pa[k]e^{-2i\pi kf\Delta t}\right|^{-2}
\end{equation}
where $P_\varepsilon$ is the constant white noise PSD. This expression is not constrained to a set of discrete frequencies as with the FFT but allows for the evaluation of the PSD at arbitrary frequencies, therefore yielding a finer representation of the Bragg peak. The combination of the AR model with the MEM for the determination of the coefficient is referred as the AR-MEM approach. If the modeled signal $s$ is non-stationary, as it is expected for any transient phenomenon, the time evolution of its frequency contents can be accounted for by allowing the AR coefficients to evolve in time \cite{book:castanie2006}. For this, successive overlapping sequences of the same length $N$ and sampling rate $\Delta t$ are processed sequentially at some other sampling rate $\tau\gg\Delta t$ and the AR-MEM coefficients are updated accordingly. Typically, a new set of AR coefficients can be obtained every $\tau=$~\SI{4}{\s} with overlapping sequences of \SI{128}{points} at the sampling rate $\Delta t=$~\SI{0.26}{\s}. An updated PSD can therefore be obtained every $\tau$~seconds as well as a new surface current estimation. We will refer to this procedure as the time-varying autoregressive modeling (TVAR) and its combination with the MEM as the TVAR-MEM.

\section{Extraction of Fast-Varying Surface Currents}\label{sec:currents}

\begin{figure}[h]
	\centering
	\includegraphics[width=\linewidth]{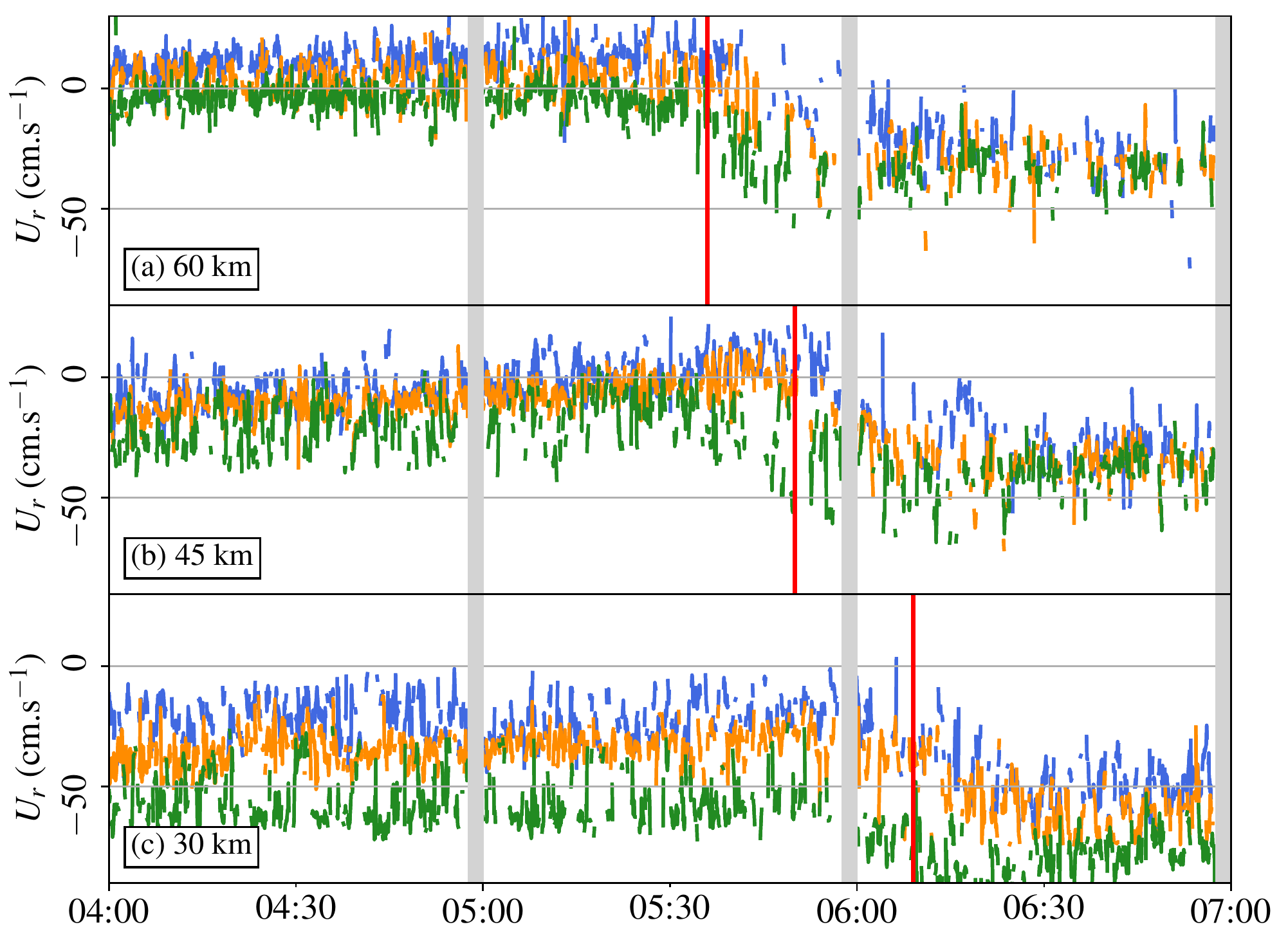}
	\caption{Time series of inverted radial surface currents $U_r$ (\si{\cm\per\s}) computed on October 14, 2016 along bearings: \ligne{RoyalBlue}{\SI{80}{\degree}}; \ligne{DarkOrange}{\SI{70}{\degree}}; \ligne{ForestGreen}{\SI{60}{\degree}} and averaged over ranges: (a) 60-\SI{63}{\kilo\metre}; (b) 45-\SI{48}{\kilo\metre}; (c) 30-\SI{33}{\kilo\metre} (see Fig. \ref{fig:tofino}). Radial currents are estimated every $\tau=$~\SI{4}{\s} using the TVAR-MEM, for overlapping intervals of $N=$~\SI{128}{points} (\SI{33}{\s}). The propagation of a jump in current magnitude is marked by \trait{red} at (a)~05:36; (b)~05:50; (c)~06:09~UTC. Gray vertical bars mark short periodic interruptions in data acquisition required by the WERA control process.}
	\label{fig:ts_meteo}
\end{figure}

The TVAR-MEM approach was used to extract the radial surface currents from the HF radar data at high temporal rate around the atmospheric event in Tofino (October 14, 05-06:00~UTC). A preliminary beam-forming operation was applied to the range-resolved backscattered complex signals recorded on each receiving antenna to resolve it in azimuth. The resulting time series was processed by blocks of \SI{33}{\s} (i.e. $N=$~\SI{128}{points}) in order to estimate the radial current every $\tau=$~\SI{4}{\s} in each range-azimuth radar cell (Fig. \ref{fig:ts_meteo}).

A synoptic view of the variations of the Doppler spectrum along different bearings can be obtained with the classical Range-Doppler maps. The horizontal axis of these maps is usually constrained by the available number of frequency bins which is limited by the integration time when a classical spectral method is employed. The use of the AR-MEM analysis allows for an increased frequency resolution and unveils fine patterns in the Range-Doppler representation which are hardly visible with a coarse FFT discretization. Fig. \ref{fig:rd_meteo} thus shows four snapshots of the obtained Range-Doppler map during the event. The radar cells are taken along the central direction, corresponding to the bearing \SI{70}{\degree} when measured from the easternmost part of the radar coverage (see Fig. \ref{fig:tofino}). The inspected region in the frequency-range domain is a high-resolution blow-up of the map around the negative Bragg line with a frequency step of \SI{1}{\milli\hertz} (as opposed to the \SI{0.03}{\hertz} resolution obtained with the FFT approach). This refined representation allows for the visualization of micro-Doppler oscillations of instantaneous Bragg lines (materialized by the ridge of maxima in blue solid lines) corresponding to fast space-time variations of the surface current around its background value averaged over one hour (black dashed lines).

\section{Synoptic View of the Event}\label{sec:cpd}

The few available operational HF radar systems for the early detection of tsunamis are based on some threshold criterion (e.g. $Q$-factor, entropy, correlation functions) indicating the probable occurrence of an abnormal, tsunami-like surface current. As mentioned, the AR method offers the possibility to monitor the latter at a high temporal rate and therefore allows for a fine estimation of the instantaneous position of the wavefront, if any. To elaborate this concept, we applied a Change Point Detection (CPD) method to obtain a systematic and automatic quantification of the space-time propagation of the wavefront. To do this, we consider individual surface current time series  $U_r(t,\rho,\theta)$ recorded in each resolved radar cell at range and azimuth position $(\rho,\theta)$; the CPD algorithm \cite{sp:truong2020} is based on segmenting the time series into $N+1$ consecutive sub-series separated by $N$ breakpoints.

The choice of the segmentation method is closely linked to the phenomenon to detect and requires a prior model. We restrained the time interval to the 3 hours surrounding the event, that is from $t_0=\textrm{04:00}$ to $t_2=\textrm{07:00 UTC}$ and modeled the phenomenon as a step function with a single break point at $t_1$. The segmentation is therefore reduced to a two-sample hypothesis testing, namely 
$\mathbb{H}_0$ before the step ($t<\hat{t}_1$) and $\mathbb{H}_1$ after the step  ($t\ge \hat{t}_1$). Here $\hat{t}_1$ is the estimated change time for each radar cell $(\rho,\theta)$, known as ``time of arrival'' (the cell index is implicit and will be omitted in the following). A wealth of segmentation methods have been proposed in the remote sensing literature. One popular method is the mean-shift approach, which has been proven to be very useful in the context of satellite imagery \cite{grsl:ming2012}. This method is based on measuring the signal empirical variances on two sliding sub-intervals of duration $\mathcal{T}$, say $V_1(t_1) = \Var[U_r(t)]$ for $t_1-\mathcal{T} \leq t<t_1$ and $V_2(t_1) = \Var[U_r(t)]$ for $t_1\leq t<t_1+\mathcal{T}$, and maximizing the change with respect to the total variance, $V_0=\Var[U_r(t)]$ for $t_1-\mathcal{T}\leq t<t_1+\mathcal{T}$. The estimated break point $\hat{t}_1$ corresponds to the time where the sum of the sub-interval variances is most different from the latter:
\begin{equation}
	\hat{t}_1 = \mathrm{arg\,max}_{t_1}\big(V_0-V_1(t_1)-V_2(t_1)\big)
\end{equation}

\begin{figure}[h]
	\centering
	\includegraphics[width=.95\linewidth]{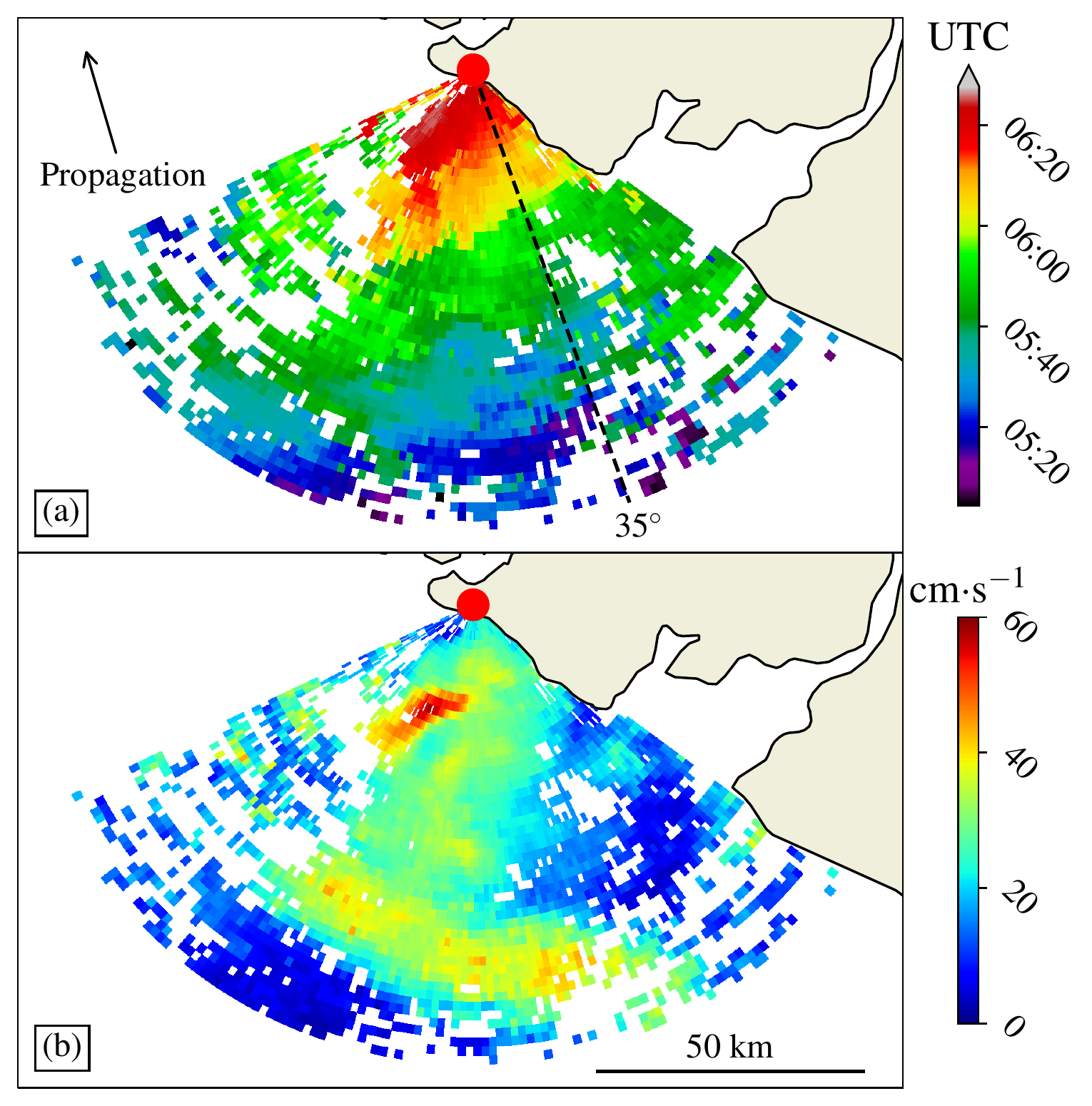}
	\caption{Results of the CPD method applied to the TVAR-MEM time series: (a) Observed arrival times (colorscale; UTC) for the October 14, 2016 event, obtained from TVAR-inverted sea surface current time series; (b)~Variation of radial surface current $\Delta U_r$ (colorscale; \si{\cm\per\s}). Arrow represents the estimated propagation direction and \dottrait{Black} represents bearing \SI{35}{\degree}.}
	\label{fig:obs_burg}
\end{figure}

A key parameter of the method is the length $\mathcal{T}$ of sliding intervals. Best detection performances result from a trade-off between a short time window and a sufficient smoothing of the oscillations due to long waves, a compromise that was found with an interval duration of about \SI{15}{\minute}. A systematic application of the CPD method to all available radar cells made it possible to map the estimated arrival time $\hat{t}_1$ of the wave front. The result is shown in Fig. \ref{fig:obs_burg}a where the different arrival times are displayed in color scales. A propagation of the phenomenon in North-West direction is clearly visible  between 05:20 and 06:00~UTC through the set of quasi-parallel color bands. Fig. \ref{fig:obs_burg}b shows the maximal amplitude of the surface current jump, $\Delta U_r$, which is defined as the change of mean between the backward and forward interval at the break point. The jump magnitude is of the order of \SI{40}{\cm\per\s} in a circular strip at about \SI{60}{\km} from radar and even reach a value of about \SI{60}{\cm\per\s} in an intense spot at closer range.

\section{Discussion}\label{sec:discussion}

\begin{figure}[h]
	\centering
	\includegraphics[width=\linewidth]{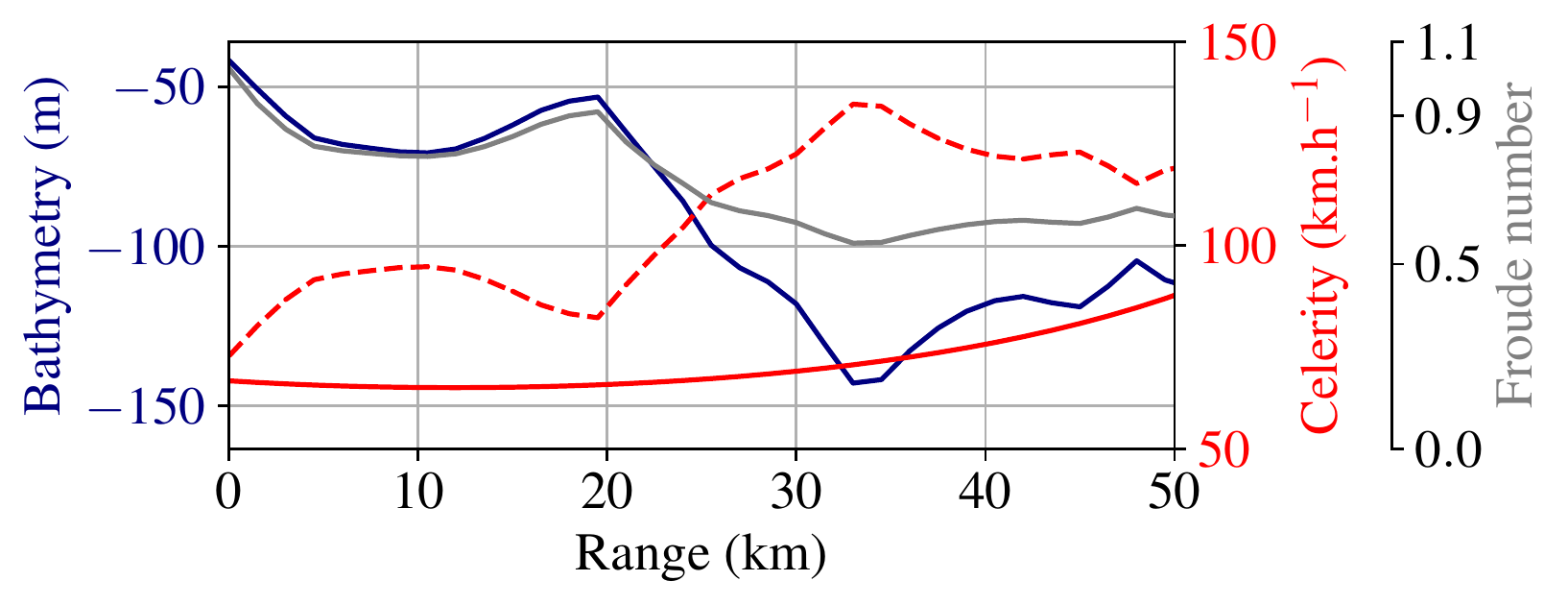}
	\caption{Bathymetry (\trait{Navy}\,; \si{\m}) and long wave celerity (\dottrait{red}\,; \si{\km\per\hour}) along the bearing \SI{35}{\degree} (see Fig. \ref{fig:obs_burg}a). The propagation speed of the current front inferred from the arrival times is also shown (\trait{red}\,; \si{\km\per\hour}) and found to be $\approx$~\SI{65}{\km\per\hour}, that is much smaller than the long wave celerity. Considering an atmospheric disturbance traveling at $U=$~\SI{75}{\km\per\hour}, Froude numbers (\trait{Grey}) are outside the ``tsunamigenic'' range $0.9<\textrm{Fr}<1.1$ \cite{nh:monserrat2006}.}
	\label{fig:speed}
\end{figure}

The combination of the TVAR-MEM and CPD approaches to process the radar signal has given an accurate synoptic view of the October 14, 2016 event and provides new insight into its geophysical origin. The event clearly belongs to the family of atmospherically induced tsunami-like sea level oscillations which can be grouped under the common denomination of ``meteotsunami'' \cite{nh:monserrat2006}. However, there are several possible amplification mechanisms of the atmospheric disturbance which can lead to the observed anomalies of residual sea level and surface currents. When the pressure variations are of the order of a few \si{\hecto\pascal}, the sea level oscillations induced by the inverse barometric effect are too small (a few \si{\cm\per\s}) to generate a visible tsunami-like wave in open sea unless they are amplified by some resonance mechanism (Proudman, Greenspan, shelf resonance, see \cite{nh:monserrat2006}). In open sea the only possible candidate for this coupling mechanism is the Proudman resonance. It requires the long wave celerity over the local bathymetry $d$ to match the atmospheric gravity wave celerity $U$, a condition which is fulfilled when the Froude number $\textrm{Fr}=U/\sqrt{gd}$ is close to $1$. A proxy for the atmospheric front propagation speed $U$ is the arrival time of the step of current which has been calculated with the combined TVAR-MEM and CPD methods in Fig. \ref{fig:obs_burg}a. By differentiating the arrival times along the main travel direction one can infer an average speed $U\approx$~\SI{65}{\km\per\hour} and local values of the Froude number which are definitely too low to excite a resonance (Fig. \ref{fig:speed}). To further evaluate the likelihood of a Proudman resonance mechanism, we investigated a possible correlation between the bathymetry and the maximum amplitude of the surface current anomaly displayed in Fig. \ref{fig:obs_burg}b. According to Green's law for shallow water gravity waves, the amplification $A$ of waves due to shoaling should scale as an inverse fourth root of depth, $A\propto d^{-\frac{1}{4}}$. Under the assumption that the observed surface current anomaly is due to the tsunami-wave orbital current, the maximum residual current amplitude should therefore also scale with the bathymetry. However, a correlation test over the radar coverage showed no systematic relationship between these two variables. This analysis leads us to disqualify a Proudman resonance to explain the sea level and current anomaly. Nevertheless, the meteorological records of the closest NOAA buoys indicate an exceptional pressure drop of the order of \SI{25}{\hecto\pascal} corresponding to a sea-level increase of \SI{25}{\cm} (see Fig. 10 from \cite{od:guerin2018}) which is consistent with the anomaly observed on the different tide gauges near Tofino.

\begin{figure}[h]
	\centering
	\includegraphics[width=.95\linewidth]{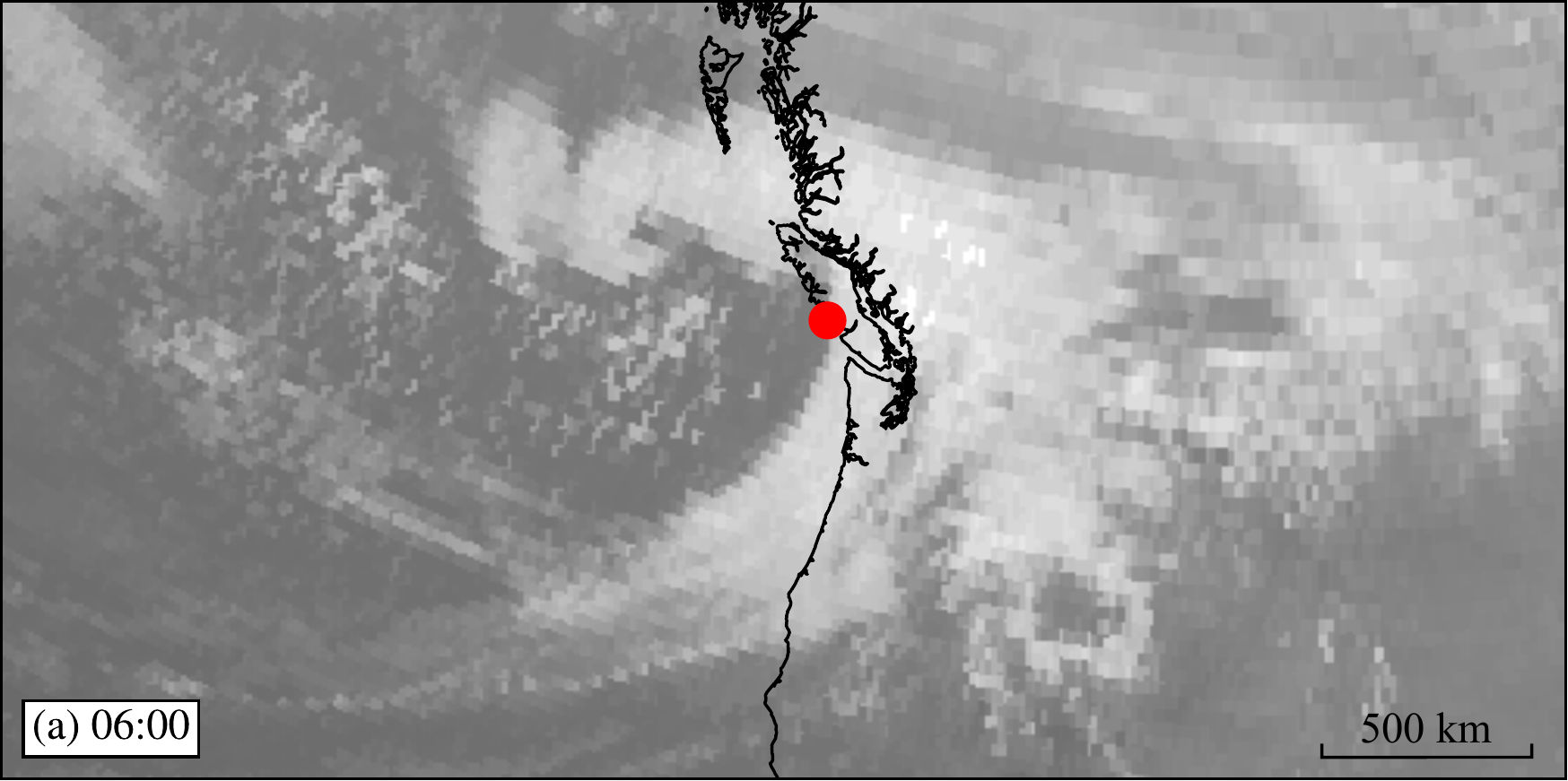}
	\includegraphics[width=.975\linewidth]{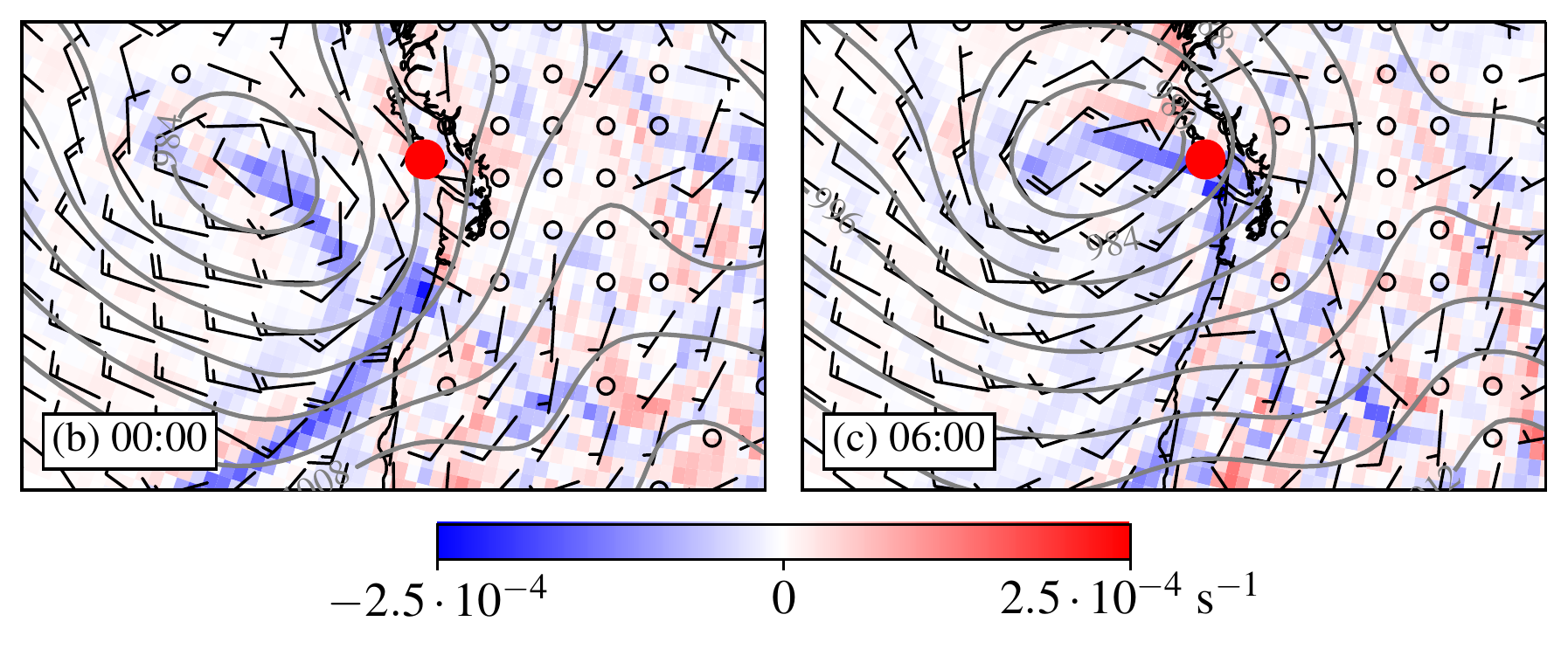}
	\caption{Synoptic views of the October 14, 2016 remnant of typhoon Songda over the Pacific Northwest:
	(a) Infrared satellite image (\SI{3.9}{\nano\m}; gray scale) captured by GOES-15 at 06:00~UTC. The clouds pattern reveals the cold front position;
	(b) and (c): Surface wind (\SI{10}{\m}; barbs; \si{\knot}) and isobars (\trait{gray}\,; each \SI{4}{hPa}) forecast by GFS at 00:00 and 06:00~UTC, resp.; along with horizontal divergence of the wind velocity field $\vb{\boldsymbol{\nabla}_H}\cdot\vb{V}$ (colorscale; \si{\per\s}).}
	\label{fig:weather}
\end{figure}

To better understand the atmospheric mesoscale source process at the origin of the event we analyzed infrared satellite images from the weather satellite GOES-15 as well as data from the GFS weather forecast model. Fig. \ref{fig:weather}a provides a synoptic view of the Pacific Northwest from GOES-15 imagery at 06:00~UTC. As seen from the cloud patterns, the eastward propagation of a low-pressure area in the remnant of typhoon Songda carried a cold front over British Columbia, which can be identified from a typical Comma feature. Figs. \ref{fig:weather}b and \ref{fig:weather}c further show the surface wind vector at \SI{10}{\m} estimated from the GFS model at 00:00 and 06:00~UTC, respectively, together with the horizontal divergence of wind velocity in colorscale. As seen, the main marked line of wind divergence matches the cold air front and reveals a pronounced low-level wind shear zone, probably turning into a squall line. Following the line of maximal negative divergence between 00:00 and 06:00~UTC, one can deduce that the atmospheric front propagates northwards at a speed of about $U\approx$~\SI{75}{\km\per\hour}, which is close to the speed inferred from the analysis of HFR residual currents. At a finer scale within the frontal zone, high-resolution models \cite{osm:rabinovich2018} show a sudden surge of strong and gusty South-Southeast winds, which are consistent with the observed travel direction. The propagating front of current is therefore likely due to a combination of Stokes drift and wind friction over the first sea surface layer, which are known to be of the order of 1-2 $\%$ of wind speed \cite{tamtare2021stokes}. The coupled analysis of HFR and spatial data thus supports the hypothesis of a mere storm surge and excludes the occurence of an actual tsunami wave.
  
\section{Conclusion}\label{sec:conclusion}

The analysis of the residual ocean current at very short integration time has allowed to identify and characterize the propagation of a low-pressure air front. This shows that oceanographic radars can be a valuable complement to spaceborne sensors for the continuous observation of strong atmospheric disturbances at a fine spatio-temporal scale. We expect that the ever increasing resolution and coverage of meteorological satellites (starting with GOES-17 as of March 2018) will allow to refine the joint space- and land-based analysis of ``tsunami-like'' events.

\section*{Acknowledgments}
\addcontentsline{toc}{section}{Acknowledgment}

The authors would like to thank Ocean Networks Canada for providing HFR data. Coastal bathymetry was provided by the British Columbia Ministry of Forests, Lands, Natural Resource Operations and Rural Development.

\bibliography{bibfile.bib}

\begin{thebibliography}{10}
\providecommand{\url}[1]{#1}
\csname url@samestyle\endcsname
\providecommand{\newblock}{\relax}
\providecommand{\bibinfo}[2]{#2}
\providecommand{\BIBentrySTDinterwordspacing}{\spaceskip=0pt\relax}
\providecommand{\BIBentryALTinterwordstretchfactor}{4}
\providecommand{\BIBentryALTinterwordspacing}{\spaceskip=\fontdimen2\font plus
\BIBentryALTinterwordstretchfactor\fontdimen3\font minus
  \fontdimen4\font\relax}
\providecommand{\BIBforeignlanguage}[2]{{%
\expandafter\ifx\csname l@#1\endcsname\relax
\typeout{** WARNING: IEEEtran.bst: No hyphenation pattern has been}%
\typeout{** loaded for the language `#1'. Using the pattern for}%
\typeout{** the default language instead.}%
\else
\language=\csname l@#1\endcsname
\fi
#2}}
\providecommand{\BIBdecl}{\relax}
\BIBdecl

\bibitem{barrick_RSE79}
D.~E. Barrick, ``{A Coastal Radar System for Tsunami Warning},'' \emph{Remote
  Sensing of Environment}, vol.~8, no.~4, pp. 353--358, 1979.

\bibitem{lipa2011japan}
B.~Lipa, D.~Barrick, S.-I. Saitoh, Y.~Ishikawa, T.~Awaji, J.~Largier, and
  N.~Garfield, ``{Japan Tsunami Current Flows Observed by {HF} Radars on Two
  Continents},'' \emph{Remote Sensing}, vol.~3, no.~8, pp. 1663--1679, 2011.

\bibitem{dzvonsko_IRS11}
A.~{Dzvonkovskaya}, D.~{Figueroa}, K.~{Gurgel}, H.~{Rohling}, and T.~{Schlick},
  ``{HF Radar Observation of a Tsunami near Chile after the Recent Great
  Earthquake in Japan},'' in \emph{12th International Radar Symposium (IRS)},
  2011, pp. 125--130.

\bibitem{lipa2012tsunami}
B.~Lipa, J.~Isaacson, B.~Nyden, and D.~Barrick, ``{Tsunami Arrival Detection
  with High Frequency ({HF}) Radar},'' \emph{Remote Sensing}, vol.~4, no.~5,
  pp. 1448--1461, 2012.

\bibitem{werachili2014}
A.~{Dzvonkovskaya}, M.~{Heron}, D.~{Figueroa}, and K.~{Gurgel}, ``{Observations
  and Theory of a Shoaling Tsunami Wave},'' in \emph{2014 Oceans - St. John's},
  2014, pp. 1--5.

\bibitem{aes:dzvonkovskaya2018}
A.~Dzvonkovskaya, ``{HF Surface Wave Radar for Tsunami Alerting: from System
  Concept and Simulations to Integration into Early Warning Systems},''
  \emph{IEEE Aerosp. Electron. Syst. Mag.}, vol.~33, no.~3, pp. 48--55, Mar.
  2018.

\bibitem{irs:dzvonkovskaya2017}
A.~Dzvonkovskaya, L.~Petersen, and T.~L. Insua, ``{Real-Time Capability of
  Meteotsunami Detection by WERA Ocean Radar System},'' in \emph{{Proc. 18th
  Int. Radar Symp.}}, Prague, Czech Republic, Jun. 2017.

\bibitem{od:guerin2018}
C.-A. Gu\'erin, S.~T. Grilli, P.~Moran, A.~R. Grilli, and T.~L. Insua,
  ``{Tsunami Detection by High-Frequency Radar in British Columbia: Performance
  Assessment of the Time-Correlation Algorithm for Synthetic and Real
  Events},'' \emph{{Ocean Dynamics}}, vol.~4, pp. 423--438, May 2018.

\bibitem{osm:rabinovich2018}
A.~B. Rabinovich, R.~Thomson, and T.~L. Insua, ``{Meteorological Tsunami of 14
  October 2016 on the Coast of British Columbia Caused by Typhoon 'Songda'},''
  in \emph{{Proc. Ocean Sci. Meeting}}, Portland, Oregon, Feb. 2018.

\bibitem{joe:domps2020}
B.~Domps, D.~Dumas, C.-A. Gu\'erin, and J.~Marmain, ``{High-Frequency Radar
  Ocean Current Mapping at Rapid Scale with Autoregressive Modeling},''
  \emph{IEEE J. Ocean. Eng.}, 2020, under revision.

\bibitem{heron_ocean15}
M.~{Heron}, A.~{Dzvonkovskaya}, and T.~{Helzel}, ``{HF Radar Optimised for
  Tsunami Monitoring},'' in \emph{OCEANS 2015 - Genova}, 2015, pp. 1--5.

\bibitem{front:roarty2019}
H.~Roarty \emph{et~al.}, ``{The Global High Frequency Radar Network},''
  \emph{{Frontiers in Marine Science}}, vol.~6, p. 164, 2019.

\bibitem{nat:crombie1955}
D.~D. Crombie, ``{Doppler Spectrum of Sea Echo at 13.56 Mc./s.}''
  \emph{Nature}, vol. 175, pp. 681--682, 1955.

\bibitem{tap:barrick1972}
D.~E. Barrick, ``{First-Order Theory and Analysis of MF/HF/VHF Scatter from the
  Sea},'' \emph{IEEE Trans. Antennas Propag.}, vol.~20, no.~1, pp. 2--10, Jan.
  1972.

\bibitem{om:guerin2018}
C.-A. Gu\'erin and S.~T. Grilli, ``{A Probabilistic Method for the Estimation
  of Ocean Surface Currents from Short Time Series of HF Radar Data},''
  \emph{{Ocean Modelling}}, vol. 121, pp. 105--116, Jan. 2018.

\bibitem{book:stoica2005}
P.~Stoica and R.~Moses, \emph{Spectral Analysis of Signals}, 2005.

\bibitem{phd:burg1975}
J.~P. Burg, ``{Maximum Spectral Analysis},'' Ph.D. dissertation, Stanford
  University, may 1975.

\bibitem{book:marple2019}
S.~L. Marple, \emph{Digital Spectral Analysis}.\hskip 1em plus 0.5em minus
  0.4em\relax Dover Publications, 2019.

\bibitem{book:castanie2006}
F.~Castani\'e \emph{et~al.}, \emph{{Spectral Analysis: Parametric and
  Non-Parametric Digital Methods}}.\hskip 1em plus 0.5em minus 0.4em\relax
  Wiley, 2006.

\bibitem{sp:truong2020}
C.~Truong, L.~Oudre, and N.~Vayatis, ``{Selective Review of Offline Change
  Point Detection Methods},'' \emph{{Signal Process.}}, vol. 167, Feb. 2020.

\bibitem{grsl:ming2012}
D.~Ming, T.~Ci, H.~Cai, L.~Li, C.~Qiao, and J.~Du, ``{Semivariogram-Based
  Spatial Bandwidth Selection for Remote Sensing Image Segmentation with
  Mean-Shift Algorithm},'' \emph{{IEEE Geosci. Remote Sens. Lett.}}, vol.~9,
  pp. 813--817, Sep. 2012.

\bibitem{nh:monserrat2006}
S.~Monserrat, I.~Vilibi{\'c}, and A.~B. Rabinovich, ``Meteotsunamis:
  atmospherically induced destructive ocean waves in the tsunami frequency
  band,'' \emph{Natural hazards and earth system sciences}, vol.~6, no.~6, pp.
  1035--1051, 2006.

\bibitem{tamtare2021stokes}
T.~Tamtare, D.~Dumont, and C.~Chavanne, ``The stokes drift in ocean surface
  drift prediction,'' \emph{Journal of Operational Oceanography}, 2021.

\end{thebibliography}

\end{document}